\documentclass[epj]{webofc}
\usepackage[utf8]{inputenc}
\usepackage[varg]{txfonts}   
\usepackage{booktabs}
\usepackage{xcolor}
\definecolor{darkred}{rgb}{0.4,0.0,0.0}
\definecolor{darkgreen}{rgb}{0.0,0.4,0.0}
\definecolor{darkblue}{rgb}{0.0,0.0,0.4}
\usepackage[bookmarks,linktocpage,colorlinks,
    linkcolor = darkred,
    urlcolor  = darkblue,
    citecolor = darkgreen]{hyperref}
\usepackage{todo}

\wocname{EPJ Web of Conferences}
\woctitle{Lattice2017}
\newcommand{\Psibar}{\overline{\Psi}}

\newcommand{\Oalpha}{\mathcal{O}(\alpha)}
\newcommand{\Oalphasquare}{\mathcal{O}(\alpha^2)}

\begin{document}
%
\selectlanguage{english}
\title{%
Isospin Breaking Corrections to the HVP with
Domain Wall Fermions

}
\author{%
\firstname{Peter} \lastname{Boyle}\inst{1}\and
\firstname{Vera}  \lastname{Guelpers}\inst{2}\fnsep\thanks{Speaker, 
\email{V.M.Guelpers@soton.ac.uk}}\and
\firstname{James} \lastname{Harrison}\inst{2}\and
\firstname{Andreas} \lastname{Juettner}\inst{2}\and
\firstname{Christoph} \lastname{Lehner}\inst{3}\and
\firstname{Antonin} \lastname{Portelli}\inst{1}\and
\firstname{Christopher} \lastname{Sachrajda}\inst{2}\\
\\
RBC/UKQCD Collaboration\\
}
\institute{%
School of Physics and Astronomy, University of 
Edinburgh, Edinburgh EH9 3JZ, United Kingdom
\and
School of Physics and Astronomy, University of 
Southampton, Southampton SO17 1BJ, United Kingdom
\and
Physics Department, Brookhaven National Laboratory, Upton, NY 
11973, USA
}
\abstract{%
We present results for the QED and strong isospin breaking corrections to the 
hadronic vacuum polarization using $N_f=2+1$ Domain Wall fermions. QED is 
included in an electro-quenched setup using two different methods,  a stochastic
and a perturbative approach. Results and statistical errors from both methods 
are directly compared with each other. 
}
\maketitle
\section{Introduction}
\label{sec:introduction}
\enlargethispage{\baselineskip}
One of the most promising quantities for the search of signatures of physics 
beyond the Standard Model at low energies is the anomalous magnetic 
moment of the muon $a_\mu$. 
A comparison of the experimental result with the Standard Model estimate 
\cite{Olive:2016xmw} yields a deviation of $\approx 3\sigma$, which might be a 
hint of new physics. However, a further investigation requires a reduction of 
the errors on the experimental and theoretical determinations of $a_\mu$.
On the theory side, the largest uncertainty arises from the hadronic vacuum 
polarization (HVP) contribution to $a_\mu$. The current most precise 
determination relies on experimental input from the cross section of 
$e^+e^-\rightarrow$hadrons \cite{Davier:2010nc,Hagiwara:2011af}. 
In the last years a lot of effort has been undertaken to calculate the HVP 
contribution to $a_\mu$ using lattice QCD (see e.g.\ 
\cite{Burger:2013jya, Bali:2015msa,Chakraborty:2016mwy, 
Borsanyi:2016lpl,DellaMorte:2017dyu}).
To be competitive with the result $e^+e^-\rightarrow$hadrons a lattice 
calculation would require an accuracy of $\lesssim1\%$. However, these lattice 
calculations are usually done in the isospin symmetric limit, i.e.\ treating 
the up and down quark as being equal. In nature there are two sources 
of isospin breaking. 
The bare masses of the up and the down quark are not 
equal and quarks have electric 
charges and thus underly QED interactions. The latter effect not only applies 
to up and down, but to all quark flavors. 
These effects are expected to be 
$\mathcal{O}((m_d-m_u)/\Lambda_\textrm{QCD})$, 
$\mathcal{O}(\alpha)\sim\mathcal{O}
(1\%)$ and thus, need to be included in a lattice calculation aiming at a 
precision of $1\%$.
\par
The structure of the proceedings is as follows: In section \ref{sec:IBlattice} 
we describe how isospin breaking and QED corrections can be included in lattice 
calculations and we show results for the HVP in section \ref{sec:results}. 
The results shown in this proceedings are already published in 
\cite{Boyle:2016lbc,Boyle:2017gzv}. Another independent calculation of QED 
corrections to the HVP for strange and charm quarks has recently be 
presented~in~\cite{Giusti:2017jof}.

\section{Isospin Breaking Corrections on the lattice}
\label{sec:IBlattice}
The expectation value of an observable $O$ can be calculated in terms of the
Euclidean path integral. The path integral including QED is given by
\begin{equation}
  \left<O\right> = \frac{1}{Z}\! \int 
\!\!\mathcal{D}[U]\,\mathcal{D}[A]\,\mathcal{D}[\Psi,\Psibar]\,\,O[\Psi,\Psibar 
,A, U]\,\,
e^ { -S_F[\Psi,\Psibar ,A, U]}\,\,e^{ - S_\gamma[A]}\,e^{ - S_G[U]}\,,
\label{eq:pathintegral}
\end{equation}
with quark fields $\Psi$, $\Psibar$, gluon fields $U$ and photon fields $A$.
The fermionic action $S_F[\Psi,\Psibar ,A, U]$ now also includes couplings of 
quarks to photons. The non-compact photon action is given by
\begin{equation}
S_{\gamma}\left[A\right]=\frac{1}{4}\sum_{x}\sum\limits_{\mu,\nu}
\left(\partial_{\mu}A_{\nu}\left(x\right)-\partial_{\nu}A_{\mu}
\left(x\right)\right)^{2}\,,\hspace{1cm}\textrm{with}\,\,\,\,\partial_\mu f(x) = 
f\left(x+\hat{\mu}\right)-f\left(x\right)\,.
\label{eq:lattice_EM_action}
\end{equation}
Two methods are commonly used to include QED in the calculation of the path 
integral \eqref{eq:pathintegral}. A stochastic method 
\cite{Duncan:1996xy} using $U(1)$ gauge configurations for the photon fields 
and a perturbative method \cite{deDivitiis:2013xla}, where the path integral 
\eqref{eq:pathintegral} is expanded in the electromagnetic coupling $\alpha$. 
The stochastic and perturbative method are explained in detail in subsections
\ref{subsec:stoch} and \ref{subsec:pert}, respectively.
\subsection{Stochastic Method}
\label{subsec:stoch}
For the stochastic method, it is convenient to use Feynman gauge, due to the 
simple structure of the photon action. 
In momentum space, the photon action in Feynman gauge is given by 
\begin{equation}
S_{\gamma,\mathrm{Feyn.}}\left[A\right]=\frac{1}{2N}\sum\limits_{k,\vec{k}
\neq0}\hat{k}^{2}\sum_{\mu}\left|\tilde{A}_{\mu}\left(k\right)\right|^{2}\,,
\label{eq:feynman_mom}
\end{equation}
where $\hat{k}=2\sin\left(k_{\mu}/2\right)$, N is the number of lattice sites 
and $\tilde{A}_{\mu}$ are the photon fields in momentum space. In this study, 
we work in QED$_L$ \cite{Hayakawa:2008an}, i.e.\ we subtract all spatial zero 
modes of the photon fields $\tilde{A}_{\mu}(k_0,\vec{k}=0)=0$.
To obtain stochastic photon fields, which are distributed according to 
$\exp(-S_{\gamma,\mathrm{Feyn.}})$, one draws Gaussian random numbers 
with variance $2N/\hat{k}^{2}$ for the components $\tilde{A}_{\mu}(k)$ of the 
photon fields in momentum space. After these photon fields have been generated 
in momentum space, we transform them to position space using a fast Fourier 
transform.
\par
In this study, we work in the electro-quenched approximation, i.e.\ we neglect 
QED effects for the see quarks. This allows to generate photon fields 
independent of the $S\!U(3)$ gluon fields. Once the photon fields are 
generated, 
they are multiplied to the gluon fields as
\begin{equation}
U_\mu(x)\rightarrow e^{-ieq_fA_\mu(x)} U_\mu(x)\,.
\label{eq:u3links}
\end{equation}
Calculations of observables then proceed as without QED but using the QCD$+$QED 
link variables \eqref{eq:u3links}. 
We average calculations using $+e$ and $-e$ to exactly remove 
$\mathcal{O}(e)$ noise \cite{Blum:2007cy}.
\subsection{Perturbative Method}
\label{subsec:pert}
\begin{figure}[h]
 \centering
 \includegraphics[width=0.9\textwidth]{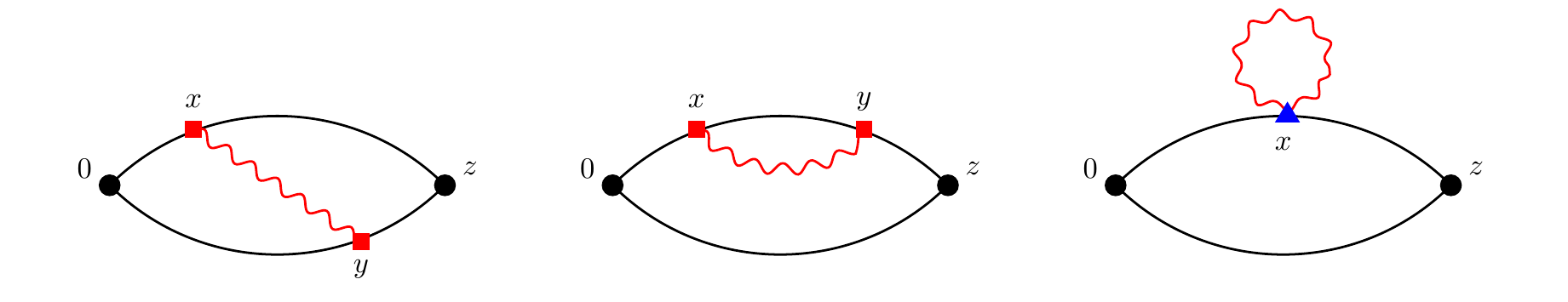}
\vspace{-0.5cm}
 \caption{From left to right: Photon exchange diagram, quark self energy 
diagram and tadpole diagram. Red squared vertices and blue triangle vertices 
denote insertions of the conserved vector current and the tadpole operator, 
respectively.}
\label{fig:diagrams}
\end{figure}

QED can be treated in a perturbative way by expanding the path integral 
\eqref{eq:pathintegral} in the electromagnetic coupling 
\cite{deDivitiis:2013xla}
\begin{equation}
 \left<O\right> = \left<O\right>_{0} + 
\frac{1}{2}\,e^2\left.\frac{\partial^2}{\partial 
e^2}\left<O\right>\right|_{e=0} + \Oalphasquare\,\,,
\label{eq:eexpansion}
\end{equation}
where the subscript $0$ indicates, that the expectation value 
$\left<\cdot\right>_0$ is calculated in pure QCD. At $\Oalpha$ one finds 
contributions from either two insertions of conserved vector currents 
$V^c_\mu$ or one insertion of the tadpole operator $T_\mu$ 
\begin{equation}
 \left<O\right> = \left<O\right>_{0} - 
\frac{(eq_f)^2}{2} \left<O
T_\mu(x)\right>_{0}\Delta_{\mu\mu}(0) 
- \frac{e^2q_fq_{f'}}{2}\left<O 
V^c_\mu(x)V^c_\nu(y)\right>_{0}\Delta_{\mu\nu}(x-y)+ \Oalphasquare\,.
\label{eq:pathintexp}
\end{equation}
In the Feynman gauge and QED$_L$ the photon propagator is given by
\begin{equation}
  \Delta_{\mu\nu}(x-y) = 
\delta_{\mu\nu}\,\frac{1}{N}\sum\limits_{k,\vec{k}\neq0}\,\,
\frac{e^{ik\cdot(x-y)}}{\hat{k}^2}\,.
\label{eq:photonprop}
\end{equation}
For the QED correction to meson masses, this gives rise to three different 
quark-connected diagrams, which are shown in figure \ref{fig:diagrams}. Here, 
we work 
in the electro-quenched setup, i.e.\ we do not consider any diagrams, where the 
photon couples to see quarks.\par
\vspace{-0.4cm}
One has to note, that additional diagrams occur, if the operator $O$ in the 
path integral itself depends on the electromagnetic coupling, as we will see 
below for the hadronic vacuum polarization.
\par
We calculate the correlation functions required for the diagrams shown in 
figure \ref{fig:diagrams} in two different ways:
\begin{itemize}
 \item \textbf{single-$\mu$ insertion:} We start by rewriting the photon 
propagator by inserting stochastic sources $\eta$
 \begin{equation}
 \Delta_{\mu\nu}(x-y) = 
\left<\sum\limits_u\Delta_{\mu\nu}
(x-u)\eta(u)\eta^\dagger(y)\right>_\eta\equiv\left<\tilde\Delta_{
\mu\nu}(x)\eta^\dagger(y)\right>_\eta\,, \hspace{1cm} 
\left<\eta(u)\eta^\dagger(y)\right>_\eta = \delta_{u,y}\,.
\label{eq:phprop_stochsources}
\end{equation}
This allows to calculate the correlation functions using sequential propagators 
with insertions of either the conserved vector current $\Gamma^c_\mu$ and 
$\tilde\Delta_{
\mu\nu}(x)$ or $\eta^\dagger(y)$. This needs to be done for each combination of 
the Lorentz indices $\mu$, $\nu$ individually and in total requires $17$ 
inversions per quark flavor and source position using Feynman gauge (where only 
diagonal terms $\mu=\nu$ contribute).
 \item \textbf{summed-$\mu$ insertion:} Here, we use four different stochastic 
sources $\xi_\mu$ one for each Lorentz index
\begin{equation}
 \Delta_{\mu\nu}(x-y) = \Big<\sum\limits_u\sum\limits_\sigma 
\Delta_{\sigma\nu}(x-u)\xi_\sigma(u)\xi^\dagger_\mu(y)\Big>_\xi\! = 
\left<\hat \Delta_\nu(x)\xi^\dagger_\mu(y)\right>_\xi\,, \hspace{0.8cm} 
\left<\xi_\sigma(u)\xi_\mu^\dagger(y)\right>_\xi\! = 
\delta_{uy}\delta_{\sigma\mu}\,,
\label{eq:hatDmunu}
\end{equation}
and calculate sequential propagators with insertions of either $\sum_\nu 
\Gamma^c_\nu \hat \Delta_\nu(x)$ or $\sum_\mu \Gamma^c_\mu 
\xi_\mu^\dagger(y)$. This requires in total $5$ inversions to construct the 
diagrams given in figure \ref{fig:diagrams}.
\end{itemize}
More details can be found in \cite{Boyle:2017gzv}.
%
\subsection{Strong Isospin Breaking}
We account for strong isospin breaking corrections in two different ways, by 
either using different bare quark masses for the valence up- and down quarks, 
such that we approximately reproduce the physical quark mass difference from 
\cite{Fodor:2016bgu} or by an expansion of the path integral in the quark 
masses as proposed in \cite{deDivitiis:2011eh}

\begin{equation}
 \left<O\right>_{m_f\neq \hat{m}} = \left<O\right>_{m_f=\hat{m}} + 
(m_{f} - \hat{m}) \left.\frac{\partial}{\partial 
m_f}\left<O\right>\right|_{m_f=\hat{m}} + \mathcal{O}\left((m_{f} - 
\hat{m})^2\right)\,,
\label{eq:mexpansion}
\end{equation}
where $\hat{m}$ is the isospin symmetric quark mass and $m_f$ the mass of 
either the up or the down quark. The derivative in \eqref{eq:mexpansion} is 
given by 
\begin{equation}
  \left.\frac{\partial}{\partial 
m_f}\left<O\right>\right|_{m_f=\hat{m}} = 
\left<O\,\mathcal{S}\right>_{m_f=\hat{m}}\,,
\end{equation}
with a scalar current $\mathcal{S} = \sum\limits_x 
\overline{\psi}_f(x)\,\psi_f(x)$.

\section{Isospin Breaking Corrections to the Hadronic Vacuum Polarization}
\label{sec:results} 
\enlargethispage{\baselineskip}
\subsection{Definitions}
\begin{figure}[h!]
 \centering
 \includegraphics[width=0.63\textwidth]{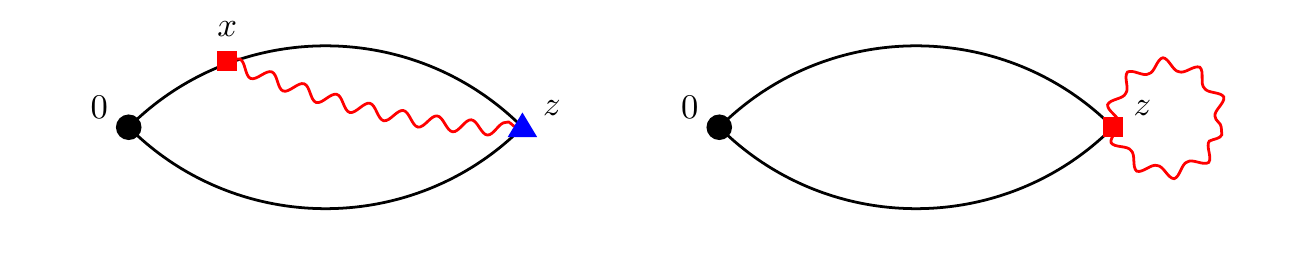}
 \vspace{-0.4cm}
 \caption{Additional diagrams for the $\Oalpha$ QED corrections to the HVP from 
expanding the conserved vector current in the conserved-local vector two-point 
function.}
\label{fig:diagrams_hvp}
\end{figure}

We calculate the hadronic vacuum polarization (HVP) tensor from a vector 
two-point 
function using a local current at the source and a conserved current at the sink
\\[-0.2cm]
  \begin{equation}
\Pi_{\mu\nu}(Q) = \sum_x e^{-iQ\cdot x} C_{\mu\nu}(x) - \sum_x C_{\mu\nu}(x) 
\,\,\,\,\,\,\,\,\,\textrm{with}\,\,\,\,\, C_{\mu\nu}(x) = Z_V\,q_f^2 \left< 
V^c_{\mu}(x) V^\ell_{\nu}(0)\right>\,,
 \label{eq:Cmunu_corr}
\end{equation}
with a zero mode subtraction \cite{Bernecker:2011gh}. We then obtain the HVP 
form factor $\Pi(\hat{Q}^2)$ from the spatial components of the HVP 
tensor\\[-0.2cm]
\begin{equation}
\Pi(\hat{Q}^2)  = \frac{1}{3} \sum_j 
\frac{\Pi_{jj}(Q)}{\hat{Q}^2}\,\,\,\,\,\,\,\,\,\,\,\textrm{with}\,\,\,\,\vec{Q}
=0\,.
\end{equation}
The QED correction to the local-conserved vector correlation function
$C_{\mu\nu}(x)$ receives two contributions. One from the QED correction to the 
two-point function itself $\delta\left< V^c_{\mu}(x) 
V^\ell_{\nu}(0)\right>$ and one from the QED correction to the multiplicative  
renormalisation $Z_V$ for the local vector current used in out setup to 
calculate the HVP
\begin{equation}
 \delta C_{\mu\nu}(x) = \delta Z_V\, q_f^2 \left< V^c_{\mu}(x) 
V^\ell_{\nu}(0)\right>_0 + Z^0_V q_f^2 \,\,\delta\left< V^c_{\mu}(x) 
V^\ell_{\nu}(0)\right>\,.
\label{eq:dCmunu}
\end{equation}
When calculating the QED correction to the vector two-point function 
$\delta\left< V^c_{\mu}(x) V^\ell_{\nu}(0)\right>$ in the perturbative method, 
one needs to take into account, that the conserved current depends on the link 
variables, and thus, when including QED, it depends on the electromagnetic 
coupling. This gives rise to two additional diagrams from the 
expansion of the conserved vector current at the sink. These diagrams are shown 
in figure \ref{fig:diagrams_hvp}.\par\vspace{-0.2cm}
\enlargethispage{\baselineskip}
\subsection{Setup of the Calculation}
For this study, we use $N_f=2+1$ dynamical flavors of Domain Wall fermions on a 
$24^3\times64$ lattice \cite{Allton:2008pn} with inverse lattice spacing of 
$1.78$~GeV. 
We use different values for the valence up- and down-quark mass, where we keep 
the up-quark mass at the mass of the sea quarks $am_u=0.005$ and increase the 
down-quark mass to $am_d=0.005915$ such that we approximately reproduce the 
physical mass difference from \cite{Fodor:2016bgu}. For the valence 
strange-quark mass we use $am_s=0.03224$ \cite{Blum:2014tka}, which corresponds 
to the physical strange quark mass without QED. The strange sea-quark mass is 
$am_s=0.04$.
The 
isospin symmetric pion mass on this ensemble is $m_\pi\approx 340$~MeV. We use 
$87$ gauge configurations and $16$ source positions for the quark propagators. 
For the stochastic method we use one $U(1)$ gauge configuration per $SU(3)$ 
gauge configuration. For the perturbative method we use one stochastic 
source $\eta$ for the single-$\mu$ 
insertion and one set of sources $\xi_\mu$ with $\mu=1,2,3,4$ for the 
summed-$\mu$ insertion per gauge configuration and source position.
%
%
%
%
%
\subsection{Results - QED corrections to the vector two-point function}
\begin{figure}[h]
\centering
  \includegraphics[width=0.4\textwidth]{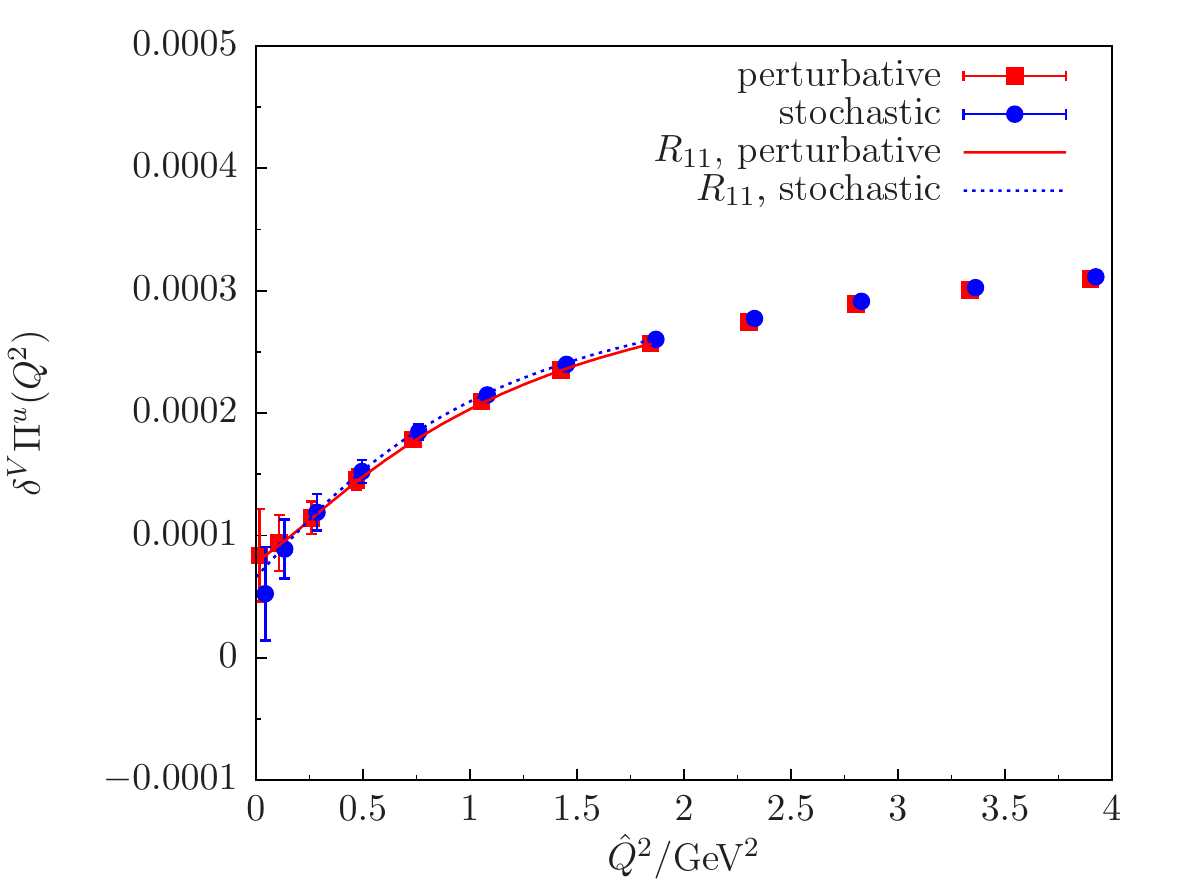}
  \includegraphics[width=0.4\textwidth]{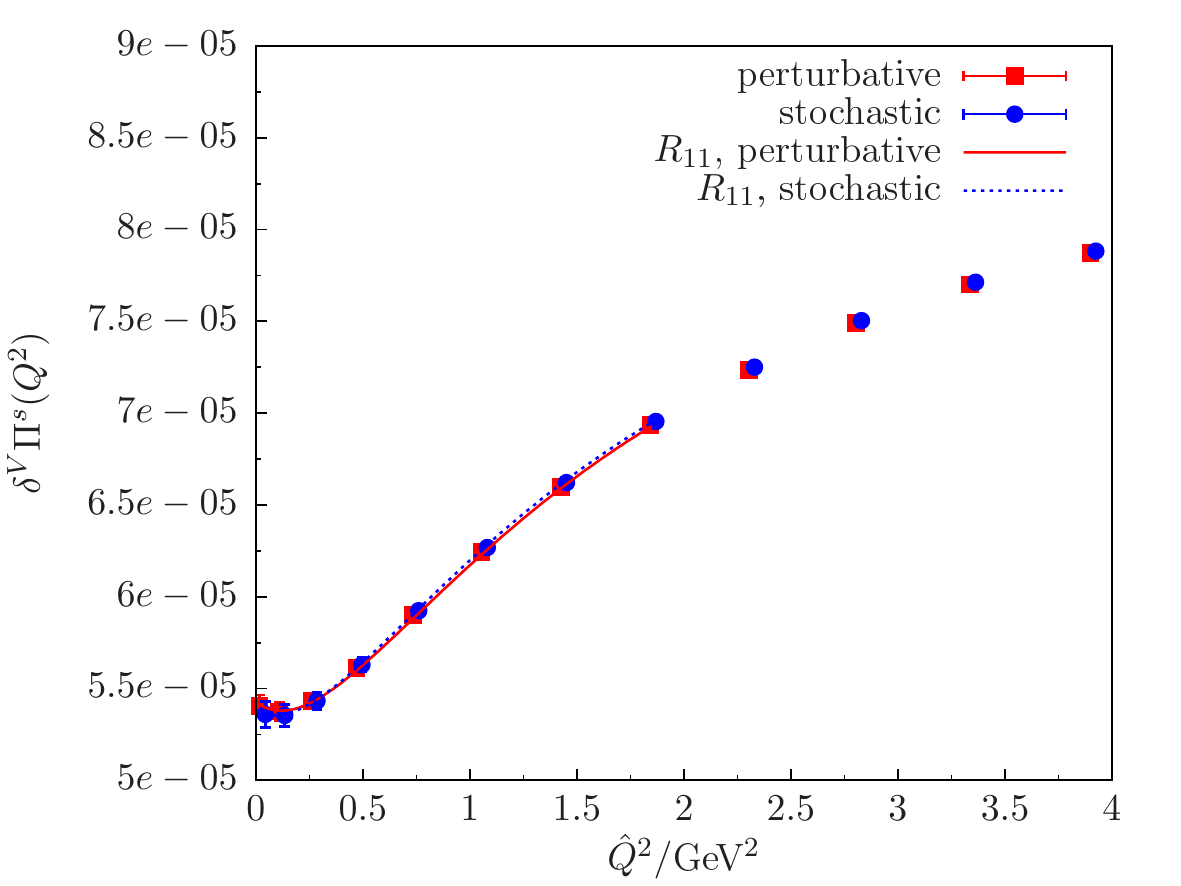}
\vspace{-0.3cm}
  \caption{QED correction to $\Pi(\hat{Q}^2)$ from the QED correction to the 
vector two-point function for the up quark (left) and the strange quark 
(right).}
\label{fig:QEDHVP}
\end{figure}

In figure \ref{fig:QEDHVP} the QED correction to the HVP from the QED 
correction to the vector two-point function is plotted against $Q^2$. Red 
squares and blue circles show results from the perturbative and the stochastic 
method respectively.
\par
To extrapolate to $Q^2=0$, we use a Pad\'e ansatz, where we allow each 
parameter in the Pad\'e to receive a QED correction
\begin{align}
 R_{11}(\hat{Q}^2) &= R^0_{11}(\hat{Q}^2) + \delta R_{11}(\hat{Q}^2) = 
\Pi^0_0 + 
\delta \Pi_0+ 
\hat{Q}^2\left(\frac{a^0+\delta a}{b^0+\delta b+\hat{Q}^2} + 
c^0+\delta c\right) 
\end{align}
 From this, we find the following fit ansatz for the $\Oalpha$ contribution
 \begin{equation}
  \delta  R_{11}(\hat{Q}^2) = \delta \Pi_0 + 
\hat{Q}^2\left(\frac{1}{b^0+\hat{Q}^2}\left[\delta  
a-\frac{\delta b\cdot a^0}{b^0+\hat{Q}^2}
\right ] + \delta c\right)\,.\label{eq:qedpade11}
 \end{equation}
The results from this fits are shown in figure \ref{fig:QEDHVP} by the solid 
red curve (perturbative data) and the dashed blue curve (stochastic data). The 
QED correction to $a_\mu$ obtained from the results of these fits, can be found 
in table \ref{tab:QEDcorr}.
\par
\begin{figure}[h]
\centering
  \includegraphics[width=0.4\textwidth]{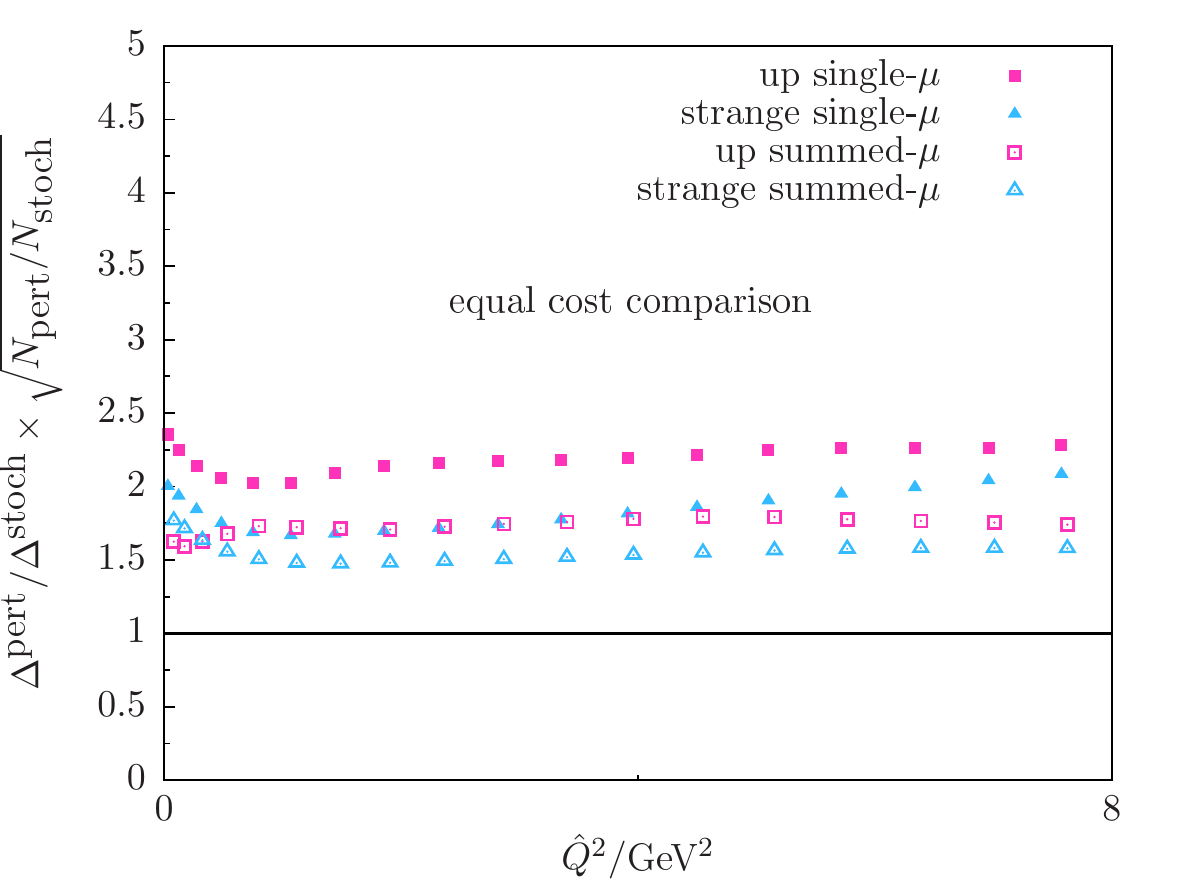}
  \vspace{-0.3cm}
 \caption{Comparison of statistical errors from the stochastic and the 
perturbative method at equal cost.}
\label{fig:HVPerr}
\end{figure}

In figure \ref{fig:HVPerr} the ratio of the statistical errors on the QED 
correction to the HVP from the perturbative over the stochastic method is 
shown. For both data sets, we have scaled the errors with 
$\sqrt{\#\textrm{inversions}}$ 
to obtain an equal cost comparison. We show results for the single-$\mu$ 
(closed symbols) and summed-$\mu$ (open symbols) insertion each for up (purple 
squares) and strange (blue triangles) quarks. We find this ratio of errors to 
be about 
$1.5-2$, indicating, the we find the statistical error from the stochastic 
method to be a factor $1.5-2$ smaller than the statistical error from the 
perturbative method at the same numerical cost.
\par\vspace{-0.2cm}
\par 
\vspace{-1cm}
\subsection{Results - QED corrections to $Z_V$}
\enlargethispage{\baselineskip}
An additional QED correction to the HVP arises from the QED correction $\delta 
Z_V$ to the multiplicative renormalisation of the local vector current. $Z_V$ 
can be determined from the ratio of the local-conserved $C^{lc}$ and the 
local-local $C^{ll}$
two-point function
\begin{equation}
 Z_V   = \frac{C^{lc}(t)}{C^{ll}(t)} = \frac{C^{lc}_0(t)+\delta 
C^{lc}(t)}{C^{ll}_0(t)+\delta 
C^{ll}(t)} = \frac{C^{lc}_0(t)}{C^{ll}_0(t)}+ 
\underbrace{\left(\frac{\delta 
C^{lc}(t)}{C^{ll}_0(t)}-\frac{C^{lc}_0(t)}{C^{ll}_0(t)}\frac
{\delta C^ { ll }(t) } { C^{ll}_0(t)}\right)}_{=\delta Z_V} + 
\Oalphasquare\label{eq:Zv0+1}\,.
\end{equation}
\par
\begin{figure}[h!]
\centering
  \includegraphics[width=0.4\textwidth]{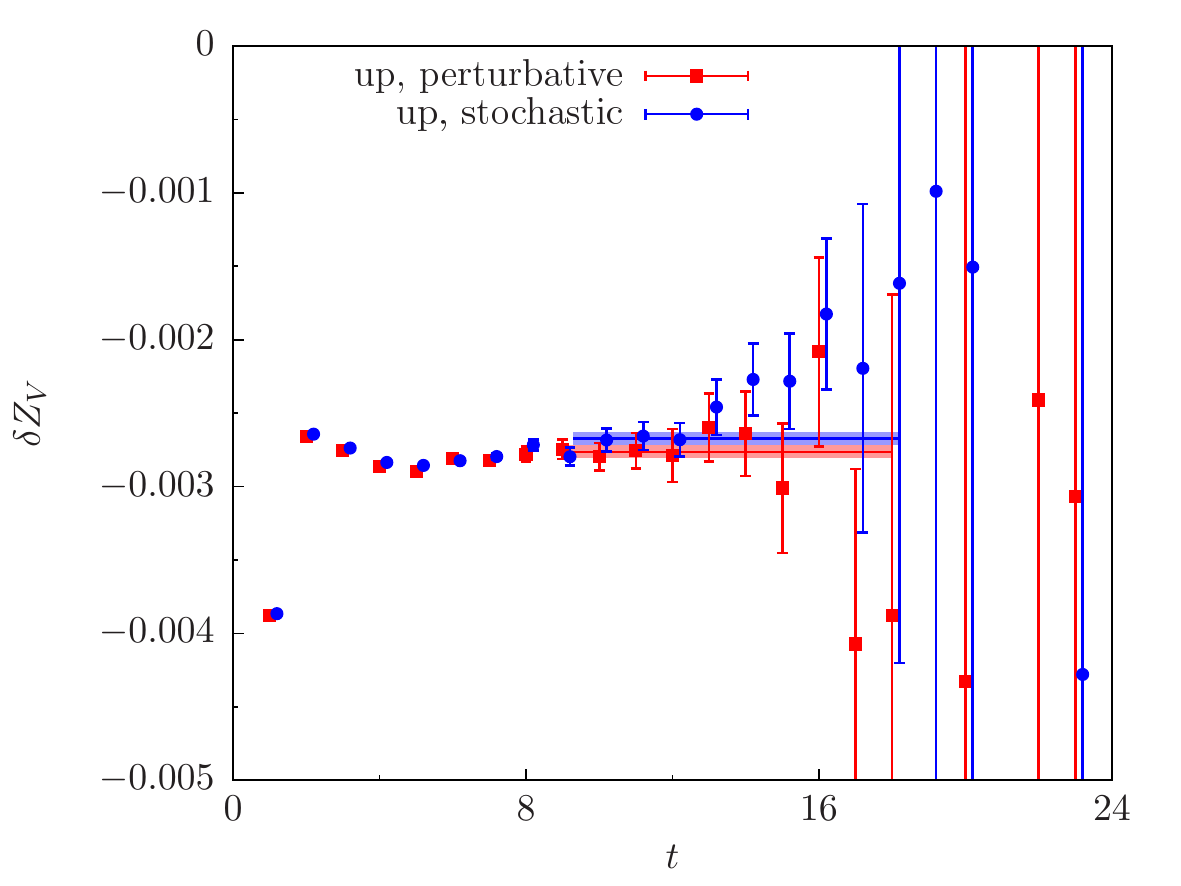}
  \includegraphics[width=0.4\textwidth]{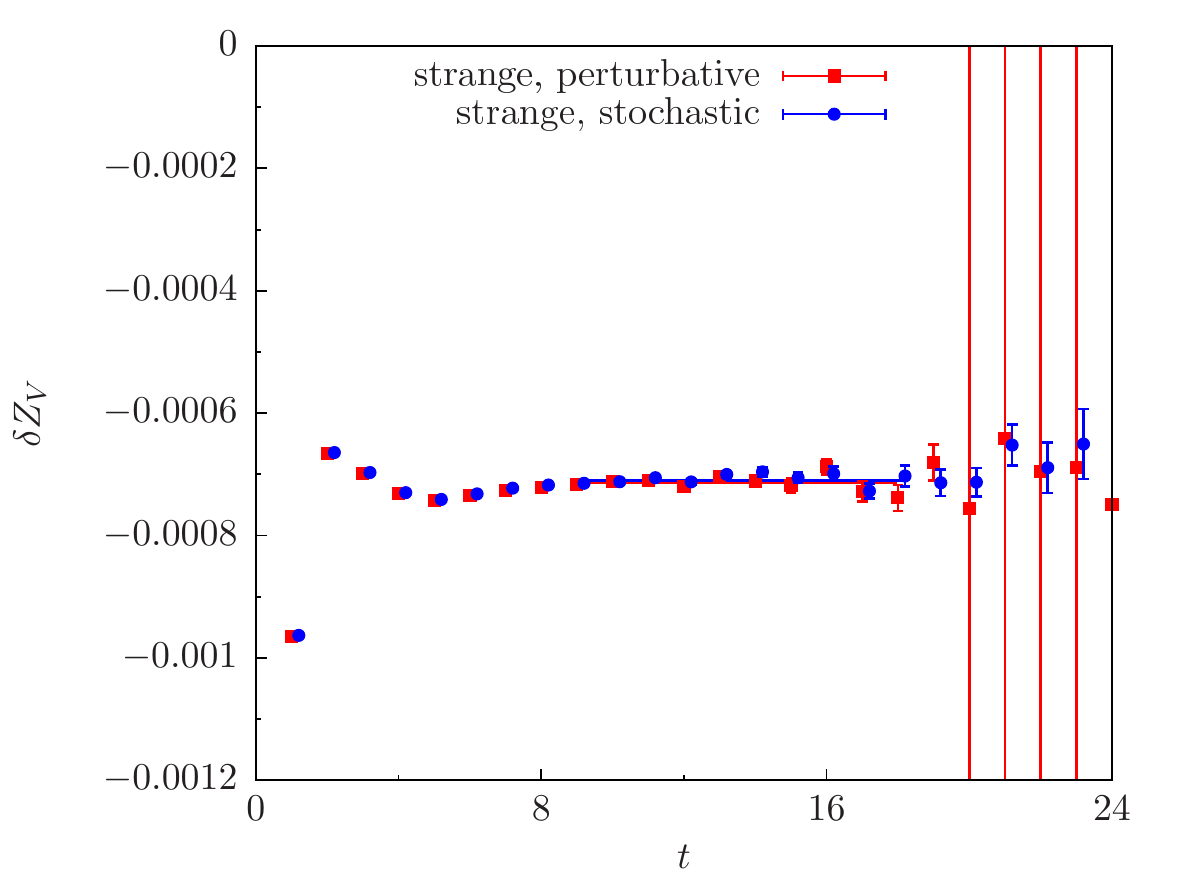}
  \vspace{-0.3cm}
 \caption{QED correction to $Z_V$ for the up quark (left) and the strange quark 
(right).}
\label{fig:dZv}
\end{figure}

\vspace{-0.2cm}
Thus, the QED correction to $Z_V$ can be obtained from a combination of 
correlation functions given in the brackets in equation \eqref{eq:Zv0+1}.
This is shown in figure \ref{fig:dZv} for the up quark (left) and the strange 
quark (right). Red square points and blue circle points denote results from the 
perturbative and the stochastic method, respectively. We have fitted a constant 
to the plateau region of the data to determine $\delta Z_V$. The results for 
the QED correction to the HVP contribution to $a_\mu$ from $\delta Z_V$ can be 
found in table \ref{tab:QEDcorr}.

\subsection{Results - strong isospin breaking corrections}
\begin{figure}[h]
\centering
  \includegraphics[width=0.4\textwidth]{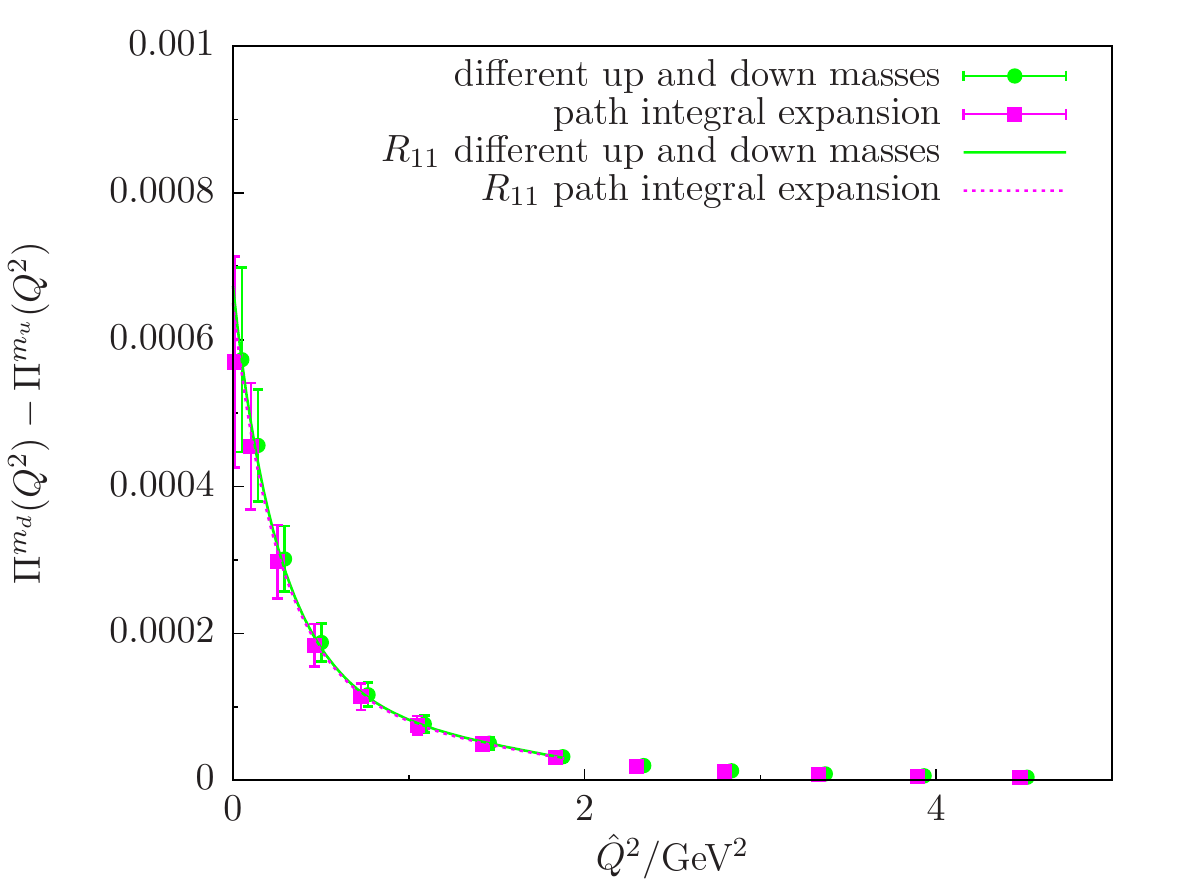}
  \vspace{-0.3cm}
 \caption{Difference of the HVP calculated with $m_u$ and $m_d$.}
 \label{fig:HVPstrong}
\end{figure}

The difference of the HVP form factor calculated with the mass of the down or 
the up-quark is shown in figure \ref{fig:HVPstrong}. Purple squares show 
results from the path integral expansion \eqref{eq:mexpansion} and green 
circles results from using different valence quark masses for the up and the 
down quark. We have fitted a Pad\'e with a similar ansatz as for the QED 
corrections to the data. The results of the fits is indicated by the curves 
shown in figure \ref{fig:HVPstrong}. We find 
$\delta_\textrm{s}a_\mu/a^u_\mu\approx-0.9\%$, where $\delta_\textrm{s}a_\mu$ 
is the difference in $a_\mu$ once calculated with the down-quark mass and once 
with the up-quark mass. In general, a lighter quark mass results in a larger 
value for $a_\mu$ and a heavier quark mass in a smaller value.
\par
\par
\vspace{-0.8cm}
\section{Summary}
Our results for the QED corrections to the HVP 
contribution to $a_\mu$ are given in table \ref{tab:QEDcorr} alongside results 
$a_\mu^0$ without QED. We give values for the two different contributions (from 
the correction to the vector-vector two-point function and from $Z_V$) 
separately.
\begin{table}[h]
\centering
\footnotesize
 \begin{tabular}{|c||c||c|c||c|c|}
 \hline
 & $a_\mu^0\times10^{10}$ & 
\multicolumn{2}{|c||}{$\delta^Va_\mu\times10^{10}$}& 
\multicolumn{2}{|c|}{$\delta^{Z_V}a_\mu\times10^{10}$}\\
 &  & stoch, $R_{11}$ 
& pert, $R_{11}$ & stoch & pert\\
\hline\hline
$u$ & $318(11)$ &  $0.65(31)$  & 
$0.37(33)$&$-1.212(52)$&$-1.249(47)$\\
$s$ & $47.98(25)$ &  $-0.0030(12)$ & $-0.0049(11)$&$-0.04886(28)$ & 
$-0.04911(27)$\\
\hline
\end{tabular}
\caption{Summary of our results for the QED correction to $a_\mu$.}
\label{tab:QEDcorr}
\end{table}
We find the QED correction to be $<1\%$ for the up quarks and $\approx0.1\%$ 
for strange quarks. Note, that these results are obtained at unphysical quark 
masses. We have not corrected our results for the QED corrections to the HVP 
by finite volume effects. We are currently studying these effects using 
lattice scalar QED \cite{James_proc}.\par
We find the strong isospin correction to be $-0.9\%$ of the isospin 
symmetric result. This result depends on our arbitrary choice
to keep the up-quark mass fixed and increase the mass of the down quark. Only a 
determination at the physical point including a tuning of the up- and down-quark 
masses to their physical values will allow to make reliable predictions for the 
actual magnitude and sign of the strong isospin breaking corrections.

\par
\vspace{0.4cm}
\textbf{Acknowledgements}\,\,\, 
This work has 
received funding from the  STFC 
Grant ST/L000296/1, the EPSRC Centre for
Doctoral Training in Next Generation Computational Modelling grant
EP/L015382/1 and
from the European Research Council under the European Union's Seventh Framework
Programme (FP7/2007-2013) / ERC Grant agreement 279757.
V.G. acknowledges partial support from the Horizon 2020 INVISIBLESPlus 
(H2020-MSCA-RISE-2015 -690575).
P.A.B. and A.P. are supported in part by UK STFC grant ST/L000458/1.
C.L. is supported in part by US DOE Contract $\#$AC-02-98CH10886(BNL) and in 
part through a DOE Office of Science Early Career Award.
This work used the DiRAC Blue Gene Q Shared Petaflop system at the
University of Edinburgh, operated by the Edinburgh Parallel Computing
Centre on behalf of the STFC DiRAC HPC Facility (www.dirac.ac.uk).
This equipment was funded by BIS National E-infrastructure capital
grant ST/K000411/1, STFC capital grant ST/H008845/1, and STFC DiRAC
Operations grants ST/K005804/1 and ST/K005790/1. DiRAC is part of
the National E-Infrastructure. Computing support for this work came partially 
from the Lawrence Livermore National Laboratory (LLNL) Institutional Computing 
Grand Challenge program.

\enlargethispage{\baselineskip}
\vspace{-0.4cm}
\bibliography{lattice2017}

\end{document}